\documentclass[twocolumn]{webofc}
\usepackage[varg]{txfonts}   
\begin{document}
\title{From cluster structures to nuclear molecules}
\title{Cluster structures, ellipsoidal shapes and nuclear molecules in light A=12-50 nuclei}

\author{
\lastname{A. V. Afanasjev}\inst{1,}\inst{2}
\fnsep\thanks{\email{Anatoli.Afanasjev@gmail.com}} 
}

\institute{Department of Physics and Astronomy, Mississippi
State University, MS 39762, USA
\and
Yukawa Institute for Theoretical Physics, Kyoto University, 606-8502 Kyoto, Japan
          }

\abstract{
  The transition from cluster structures to extremely elongated ellipsoidal shapes and nuclear molecules in light $A=12-50$ $(N \sim Z)$ nuclei has been studied within the framework of 
covariant density functional theory. Nodal structure of the occupied single-particle states
plays a critical role in microscopic  understanding of this transition. This is illustrated by 
the analysis of dominant types of single-particle density distributions and their evolution
(from the bottom of nucleonic potential) with deformation and particle number. 
The  microscopic mechanism of the transition from clustered structures to ellipsoidal shapes 
and  nuclear molecules and between them is  discussed.}
\maketitle
%

\section{Introduction}
\label{intro}

  Light nuclei are characterized by the variety of the types of nucleonic configurations  
with different nuclear shapes. Ellipsoidal shapes are the most abundant ones, but there 
are also cluster structures and nuclear molecules.  Cluster and extremely deformed 
structures in light nuclei have attracted a considerable interest  (both experimental and  theoretical) in recent years (see Refs.\ 
\cite{40Ar.10,K.12,EKNV.12,S-34-mol.14, IIIMO.15,KSK.16,RA.16,FHHKLM.18}). Different models have been applied to the 
description of such structures. Cluster models provide an important insight into cluster 
dynamics of nucleus but are limited by initial assumptions about clusters and impossibility 
to describe many shell model configurations. Note that the cluster description does not 
correspond to clearly separated  $\alpha$-particles, but generates the mean-field states 
largely by antisymmetrization  \cite{MKKRHT.06}.
 
  Density functional theory (DFT) is alternative framework which has been applied to the
description of the variety of nuclear structures in light nuclei 
\cite{ER.04,ASPG.05,RMUO.11,EKNV.12,EKNV.14,YIM.14,RA.16,AA.18}.
It does not assume the existence of cluster structures and allows simultaneous treatment 
of cluster and mean-field-type states.  In this framework, the formation of clusters proceeds 
from microscopic single-nucleon degrees of freedom via many-body correlations. 

  In this manuscript, a short review of recent results obtained for light $A=12-50$
nuclei  within the covariant density functional theory (CDFT \cite{VALR.05}) is
presented\footnote{ For results obtained in non-DFT frameworks for light $A\leq 20$
nuclei I refer the readers 
to recent extensive review \cite{FHHKLM.18}.}. A special attention is paid to the 
discussion of the microscopic mechanism of the transition between different nuclear 
shapes.

\section{Nuclear shapes in the $N\sim Z$ $A=12-50$ mass region.}
\label{nuc_shapes}

\begin{figure*}[htb]
\centering
\vspace*{0cm} 
\includegraphics[width=17.2cm,clip]{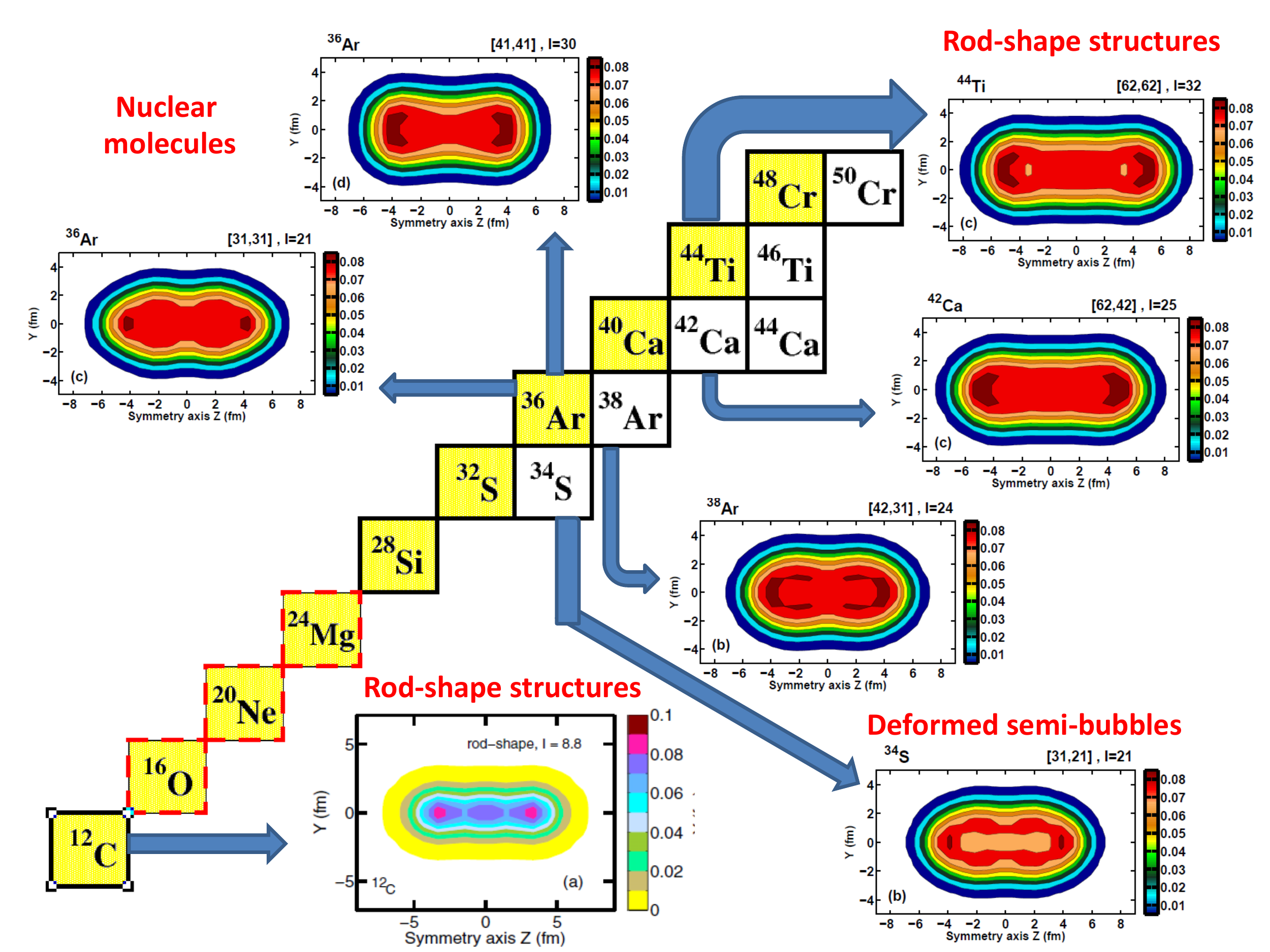}
\caption{The nuclei in which extremely deformed and rod-shaped structures have been 
studied in the DFT calculations during last decade. The boxes with black lines are used 
for the nuclei which have been studied in the CDFT framework by us in Refs.\ 
\cite{RA.16,AR.16,AA.18,AAA.18}. Yellow background is used for the nuclei with $N=Z$. 
Note that rod-shaped structures in the $^{16}$O  and $^{24}$Mg nuclei 
(indicated by boxes with red dashed lines) and in the $^{12}$C nucleus have been studied  
in the covariant and Skyrme DFT frameworks in Refs.\ \cite{IMIO.11,YIM.14,ZIM.15,IIIMO.15}. 
Neutron density distributions of several nuclei are plotted in order to indicate the variety  of 
reflection symmetric nuclear shapes possible in this mass region (see text for details). Based 
on the results presented in Refs.\ \cite{RA.16,AR.16,AA.18,AAA.18}.  Note that the density 
plot for $^{12}$C is taken from Ref.\ \cite{AA.18}: as compared with other density plots  it 
has different colormap but keeps the units of length in the $y$ and $z$ directions consistent 
with other density plots. The density plots provide also information on the configurations of these
states and spin values at which they were calculated (see Refs.\ \cite{AR.16,AA.18}) for more
details).}
\label{fig-chart}
\end{figure*}

 Over recent years a number of the studies of the cluster structures,
ellipsoidal shapes and nuclear molecules has been undertaken both at
spin zero and at high spins in the CDFT framework 
\cite{ASPG.05,EKNV.12,EKNV.14,ZIM.15,YIM.14,AR.16,RA.16,AAA.18,AA.18}.
The results for ground states presented in Refs.\  \cite{ASPG.05,EKNV.12,EKNV.14} 
are not discussed here since extremely deformed shapes which become 
yrast or near yrast at high spin are in the focus of the present manuscript.
Fig.\ \ref{fig-chart} shows both the 
part of nuclear chart in which exotic shapes have been studied at high spin 
in the CDFT framework and most interesting examples of exotic nuclear 
shapes. These studies  have been performed using cranked relativistic mean 
field (CRMF) approach \cite{KR.89,AKR.96}  in which the pairing correlations 
are neglected. The calculated configurations are labeled by shorthand 
[$n_{1}$($n_{2}$)($n_{3}$),$p_{1}$($p_{2}$)($p_{3}$)] labels \cite{RA.16}, 
where $n_{1}$, $n_{2}$ and $n_{3}$ are the number of neutrons 
in the $N=3$, 4 and 5 intruder/hyperintruder/megaintruder orbitals 
and $p_{1}$, $p_{2}$ and $p_{3}$ are the number of protons in the 
$N=3$, 4 and 5 intruder/hyperintruder/megaintruder orbitals. The labels 
$n_{i}$ and $p_{i}$ are omitted in the labeling of the configurations if 
the respective orbitals are not occupied. Note that proton and neutron total 
and single-particle densities are extremely similar because of the 
$N\approx Z$ character of the nuclei under study \cite{AA.18}.  In addition, 
calculated  configurations of interest have nearly axial shapes 
\cite{ZIM.15,YIM.14,RA.16,AR.16}.

 A linear chain of three $\alpha$ clusters, leading to ``rod-shaped'' nucleus,
is calculated in $^{12}$C in Refs.\ \cite{EKNV.14,ZIM.15}.  Its density 
distribution is shown  in left bottom corner of Fig.\ \ref{fig-chart}. A number
of similar $\alpha$-cluster configurations has been found in neighboring
C nuclei (see Refs.\ \cite{ZIM.15,EKNV.17}.)  Another example of rod-shaped
nucleus is linear chain configuration of four $\alpha$-clusters in $^{16}$O
investigated using Skyrme cranked Hartree-Fock method in Ref.\  
\cite{IMIO.11} and CRMF approach in Ref.\ \cite{YIM.14}. However, as shown
in Fig.\ \ref{fig-chart} with increasing mass number the formation of the $\alpha$-cluster
structures becomes significantly less likely and at extreme deformations there is
a competition of the nucleonic configurations with ellipsoidal (see Ref.\ 
\cite{RA.16}) and  rod-shaped  (the [62,62] configuration of $^{44}$Ti and
the [62,42] configuration of $^{42}$Ca in  Fig.\ \ref{fig-chart}) shapes.
Moreover, more exotic shapes such as nuclear molecules (the [31,31] and 
[41,41] configurations of  $^{36}$Ar in Fig.\ \ref{fig-chart}) and deformed 
semi-bubbles (the [31,21] configuration of $^{34}$S in Fig.\ \ref{fig-chart}) are 
formed. A specific  feature of the nuclear molecules is the existence of two fragments 
connected by the neck while deformed semi-bubbles are characterized by the
central (elongated in shape) depression in the density distribution. It is clear
that there should be some microscopic mechanism which dictates the transition
from one type of shape to another one and the evolution/dominance of available 
shapes as a function of particle numbers. This mechanism is discussed in next 
section.

\begin{figure*}[ht]
\centering
\vspace*{0cm}
\includegraphics[width=17.2cm,clip]{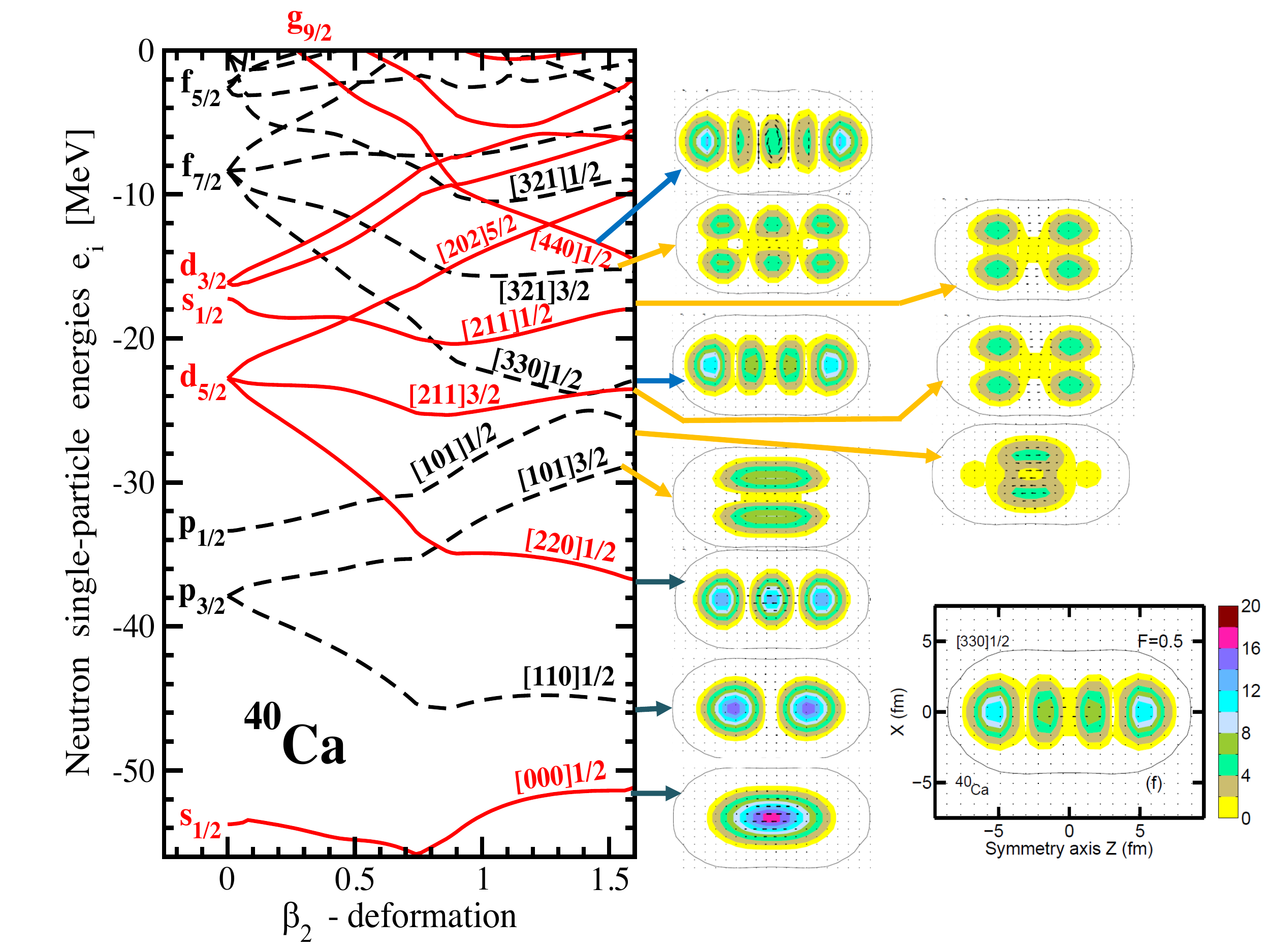}
\caption{(left panel) Single-particle energies, i.e., the diagonal elements of the single-particle
Hamiltonian $h$ in the canonical basis \cite{RS.80}, for the lowest in total energy solution in the 
nucleus $^{40}$Ca calculated as a function of the quadrupole equilibrium deformation $\beta_2$. 
The upper value of $\beta_2$  approximately corresponds to quadrupole deformations typical 
for  megadeformed configurations in this nucleus. The calculations are performed in axial 
relativistic Hartree-Bogoliubov framework \cite{AARR.14} using the NL3* covariant energy 
density functional \cite{NL3*}. The orbitals are labelled by means of asymptotic quantum 
numbers (Nilsson labels) $[Nn_z\Lambda]\Omega$. (right panels)  Single-neutron density 
distributions due to  the occupation of indicated Nilsson states. The box in right bottom 
corner exemplifies physical dimensions of the nucleus as well as the colormap used for 
single-particle densities. Other density plots are reduced down to the shape and size of the 
nucleus which is indicated by black solid line corresponding to total neutron density line of 
$\rho=0.001$ fm$^{-3}$. The colormap shows the densities as multiplies of $0.001$ fm$^{-3}$; 
the plotting of the densities starts  with yellow color at $0.001$ fm$^{-3}$. The densities are 
based on the results of Ref.\  \cite{AA.18}. 
}
\label{fig-Nilsson}
\end{figure*}

\begin{figure*}[ht]
\centering
\vspace*{0cm} 
\includegraphics[width=16.8cm,clip]{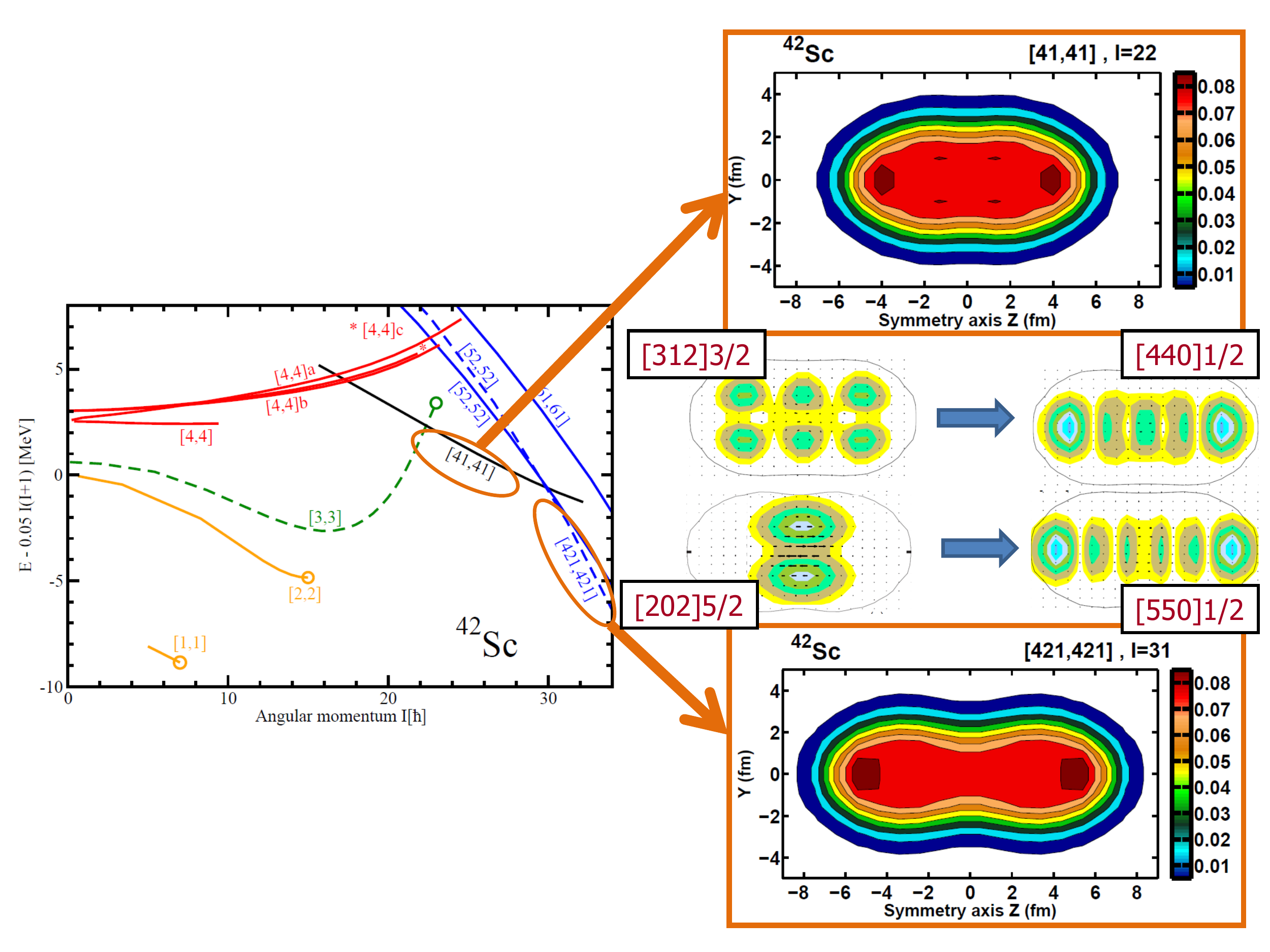}
\caption{(left panel) Energies of the calculated configurations in $^{42}$Sc relative 
to a smooth liquid drop reference $AI(I+1)$, with the inertia parameter $A=0.05$. 
(right column, top and bottom panels). The self-consistent proton density $\rho_p(y,z)$ 
as a function of $y$ and $z$ coordinates for the indicated configurations at specified 
spin  values. The equidensity lines are shown  in steps of 0.01 fm$^{-3}$ starting from 
$\rho_p(y,z)=0.01$ fm$^{-3}$.  (right column, middle) The single-particle densities
of the orbitals by which the [41,41] and [421,421] configurations differ. The direction
of particle-hole excitations by which the [421,421] configuration is created from the
[41,41] one (see text for more details) is shown by large dark blue arrows. Based on 
the results of Refs.\  \cite{RA.16,AA.18}.}
\label{fig-molecule}
\end{figure*}

\section{Underlying single-particle structure and the role of nodal structure 
of the single-particle wave functions}
\label{sp_structure}

  Many features of light nuclei depend sensitively on the underlying single-particle 
structure. The wave function $\Psi_{[Nn_z\Lambda]\Omega}$ of the single-particle 
state denoted by the Nilsson quantum number $[Nn_z\Lambda]\Omega$ can be 
expanded  into the basis states $|N'n'_z\Lambda'\Omega'>$ by 
\begin{eqnarray}
\Psi_{[Nn_z\Lambda]\Omega} = \sum_{N'n'_z\Lambda'\Omega'} c_{N'n'_z\Lambda'\Omega'} |N'n'_z\Lambda'\Omega'>
\label{wave-funct}
\end{eqnarray}
which leads to the single-particle density $\rho_{[Nn_z\Lambda]\Omega}$ given by
\begin{eqnarray}
\hspace{-0.8cm}\rho_{[Nn_z\Lambda]\Omega} = \sum_{N'n'_z\Lambda'\Omega'}  c^2_{N'n'_z\Lambda'\Omega'}  
<N'n'_z\Lambda'\Omega'|N'n'_z\Lambda'\Omega'>  
\end{eqnarray}
 Here, the basis states are characterized by principal quantum number $N'$, 
the number $n'_z$ of nodes in the axial direction 
($z$-direction)  and the projections of orbital ($\Lambda'$) and total 
($\Omega'$) single-particle angular momenta on the axis of symmetry. The 
sum in these equations runs over all allowable combinations of the 
quantum numbers $N', n'_z, \Lambda'$ and $\Omega'$.

  The weights  $c^2_{N'n'_z\Lambda'\Omega'}$ define the contributions of the 
basic states $|N'n'_z\Lambda'\Omega'>$ into the single-particle density. Table
\ref{table-wf} shows that in the nuclei/deformations of interest the single-particle 
wave functions are dominated by a  single very large component (basis state) 
which in turn will  define the spatial distribution of the single-particle density.
This domination is especially pronounced for the single-particle orbitals located 
at the bottom of the nucleonic potential; the rotation somewhat reduces this 
dominance but still preserves it (see Ref.\ \cite{AA.18}).

These features are extremely important for an understanding of the $\alpha$-clusterization
and the evolution of extremely elongated shapes in light nuclei since these basis states 
define the nodal structure of the wavefunctions of the single-particle  states and their 
density distributions. Three basic types of single-particle density distributions, namely, 
spheroidal/elipsoidal  shapes, doughnut and ring shapes, play an important role in 
forming the nuclear shapes at large elongation.  In addition, there is very simple 
connection between the type of single-particle density distribution and the Nilsson label
of the state \cite{AA.18}:

\begin{itemize}
\item
  Axial or nearly axially symmetric spheroidal/elipsoidal like density clusters are formed 
by the Nilsson states with the $[NN0]1/2$  quantum numbers. This is because dominant 
basis states of these Nilsson states have no nodes in radial direction and $n_z=N$ nodes 
in axial direction.  Density clusters with highest densities are located in polar regions 
for $N>0$.

\item
Doughnut density distributions are formed by  the Nilsson $[N01]\Omega$ states 
because the densities of their dominant basis states have one node in radial 
direction and no nodes in axial direction.

\item
Multiply (two for $n_z=1$ and three for $n_z=2$) ring shapes for $N=2$ and 3
are formed by Nilsson states with the structure $[N,N-1,1]\Omega$.
\end{itemize}

  These features are the consequences of the dominance of the wavefunction by
a single basis state where only two types of the basis states are important, namely, 
the $|N,N,0>$ and $|N,N-1,1>$ states. The density distributions of the basis states 
with $|NN0>$ will be axially symmetric with the maximum of density located at the 
axis of symmetry. The basis states with  $|N,N-1,1>$ structure will have a zero 
density at the axis of symmetry.

  Fig.\ \ref{fig-Nilsson} shows the evolution of the energies of the single-particle
states with quadrupole deformation in the $^{40}$Ca nucleus and the single-particle 
densities of these states at the deformations typical for megadeformed configurations 
in  this nucleus. One can see that these density distributions show all the
features discussed above. In addition, this figure clearly indicates the importance
of the deformation which has two critical effects. First, it leads to the formation of
density clusters with specific nodal structure [remember, the spherical states from 
which deformed states originate have spherical density distributions] and to the
separation of the clusters in space. This is especially critical for the formation of
the $\alpha$-clusters. 

   Second, it lowers the energies of the Nilsson states of the $[NN0]1/2$ type 
which favor the $\alpha$-clusterization and leads to the situation when all occupied 
Nilsson states have this type of structure.  Indeed, the occupation 
of the [000]1/2, [110]1/2 and [220]1/2 proton and neutron orbitals is responsible for 
triple $\alpha$-cluster structure in $^{12}$C (see Fig.\ \ref{fig-chart} and Refs.\ 
\cite{ZIM.15, AA.18}).  Interestingly enough the occupation of these states in the 
megadeformed [42,42] configuration of $^{40}$Ca still forms  triple $\alpha$-cluster 
structure (see Ref.\ \cite{AA.18}). However, the total wavefunction of this configuration 
is more complex due to the presence of other single-particle orbitals which do not favor 
the formation of $\alpha$-cluster structures. The formation of the chain of four 
$\alpha$-clusters will require the occupation of four lowest states of the $[NN0]1/2$ 
type.  Fig.\ \ref{fig-Nilsson} clearly shows that such a situation is achievable by modest 
increase of quadrupole deformation beyond $\beta_2=1.6$ leading to the lowering of 
the  [330]1/2 state below the [101]3/2 state. Such structure has been studied
in the $^{16}$O nucleus in the Skyrme DFT  and CDFT in Refs.\ \cite{IMIO.11,YIM.14}.  
However, the formation of the chain of five $\alpha$-clusters is more energetically expensive since 
it requires substantial increase of quadrupole deformation leading to the lowering of 
the [440]1/2 state below the [321]3/2, [211]1/2, [211]3/2, [101]1/2 and [101]3/2 states 
(see Fig.\ \ref{fig-Nilsson}). Even higher energy cost and higher deformation
are required for the formation of the chain of 6 $\alpha$-clusters.

  It is necessary to recognize that these $\alpha$-cluster chains are built by 
multiply particle - multiply hole excitations with respect of the ground state 
configuration.   As a result, they are located at high excitation energies (see Refs.\  
\cite{ZIM.15,AA.18,IMIO.11,YIM.14}) and thus are very difficult to observe in 
experiment. The low-energy excited states in the mass region of interest are 
build at either spherical or normal/high deformation or superdeformation. Fig.\ 
\ref{fig-Nilsson}  clearly indicates that at these deformations nucleonic configurations 
are typically build by the mixture of the $|N,N,0>$ and $|N,N-1,1>$ states and the 
latter states become more abundant with the increase of particle number. Moreover,
at low deformations the nodal structure of the single-particle density distributions becomes
affected by increased mixing between the basis states and the separation between 
the density clusters becomes reduced so that separate $\alpha$-clusters could
not be formed. As a result, apart of well-know two $\alpha$-cluster structure in 
$^{8}$Be and similar structures in neighboring nuclei \cite{OFE.06}, the possibilities for 
the formation of $\alpha$-clusters at such deformations are substantially reduced. 
In reality, calculated shapes of absolute majority of nucleonic configurations at 
such deformations are ellipsoidal.
 
  The situation is similar at megadeformation ($\beta_2 \sim 1.6$) for particle 
numbers above 6 (see Fig.\ \ref{fig-Nilsson}).  The  $|N,N-1,1>$ states  with 
doughnut or mutitply-ring single-particle densities  become more abundant than 
the $|N,N,0>$ states. As a result, the contribution of the states, favoring 
$\alpha$-clusterization, into the structure of the total wavefunction decreases
and ellipsoidal nuclear shapes as well as nuclear molecules become dominant
species of nuclear shapes at these particle numbers.

\begin{table}[ht]
\centering
\caption{The squared amplitudes $c^2_{N'n'_z\Lambda'\Omega'}$ of two largest 
components of the wave functions of the neutron single-particle states occupied in 
the megadeformed [42,42] configuration of $^{40}$Ca at spin zero. The states are 
shown from the bottom of nucleonic potential in the same sequence as they appear 
in the routhian diagram of  Fig.\ 2b in Ref.\  \cite{AA.18} at frequency $\Omega_x=0.0$ MeV.  
\label{table-wf}
}
\begin{tabular}{|c|c|c|} \hline 
  State    &             Wave function                     \\ \hline
$[000]1/2$ &       92.7\%$|000,1/2>$ + 6.7\%$|220,1/2>$ +    \\
$[110]1/2$ &       89.6\%$|110,1/2>$ + 9.6\%$|330,1/2>$ +   \\
$[220]1/2$ &       85.1\%$|220,1/2>$ + 6.5\%$|000,1/2>$ +    \\
$[101]3/2$ &        92.9\%$|101,3/2>$ + 6.7\%$|321,3/2>$ +   \\
$[330]1/2$ &        75.7\%$|330,1/2>$ + 8.5\%$|110,1/2>$ +    \\
$[211]3/2$ &        92.5\%$|211,3/2>$ + 6.9\%$|431,3/2>$ +    \\
$[101]1/2$ &       90.2\%$|101,1/2>$ + 5.0\%$|330,1/2>$ +    \\
$[211]1/2$ &       81.4\%$|211,1/2>$ + 9.6\%$|431,1/2>$ +    \\
$[321]3/2$ &       85.7\%$|321,3/2>$ + 6.1\%$|101,3/2>$ +    \\
$[440]1/2$ &       77.4\%$|440,1/2>$ + 10.7\%$|211,1/2>$ +    \\
$[550]1/2$  &       79.3\%$|550,1/2>$ + 6.8\%$|541,1/2>$ +    \\ \hline
\end{tabular}
\end{table}

\section{Transition from ellipsoidal shapes to  nuclear molecules by means of particle-hole
                             excitations}
\label{nucl_molec}

  The results of systematic investigations of nuclear shapes at high spin performed in 
Ref.\ \cite{RA.16} show interesting examples of the coexistence of elipsoidally shaped 
structures and nuclear molecules in the same nucleus at comparable spins and frequently  
at similar elongations of nuclear shape. It turns out that the configurations of these two 
types of the shapes are connected by characteristic particle-hole excitations (see Ref.\ 
\cite{AA.18}).  In order to build nuclear molecules from typical ellipsoidal density distributions 
one has to move the matter from the neck (equatorial) region into the polar regions of the 
nucleus. This can be achieved by means of specific particle-hole excitations which move 
the particles from (preferentially) doughnut type orbitals or from the orbitals which have 
a density ring in a equatorial plane into the orbitals (preferentially of the $[NN0]1/2$ type) 
building the density mostly in the  polar regions of the nucleus.

  This microscopic mechanism of the creation of nuclear molecule is illustrated in Fig.\ 
\ref{fig-molecule} on the example of the hyperdeformed (HD) [41,41] and megadeformed (MD) 
[421,421] configurations in the $^{42}$Sc nucleus. The [41,41]  configuration is yrast in the spin 
range $I \approx 23-30\hbar$ and it has ellipsoidal shape. At spin $I\sim 30\hbar$, it is 
crossed by the MD [421,421] configuration which is yrast at $I\geq 30\hbar$ (see left panel 
in Fig.\ \ref{fig-molecule}). This configuration is nice example of nuclear molecule in which 
two fragments are connected by the neck. This transition is achieved in proton and neutron subsystems 
by the particle-hole excitations from the $[202]5/2$ and $[321]3/2$ orbitals into the 
$[440]1/2(r=+i)$ and $[550]1/2(r=+i)$ orbitals. At the deformations and rotational frequencies 
of interest, these orbitals are located in the vicinity of each other (see Figs. 3 and 4 in Ref.\ 
\cite{RA.16}) and this explains why the total energies of these two configurations are 
comparable (see left panel in Fig.\ \ref{fig-molecule}). The  $[202]5/2$ orbital has doughnut 
type density distribution and $[321]3/2$  has triple ring density distribution (see middle
of right column in Fig.\ \ref{fig-molecule}); these two orbitals are occupied in the HD [41,41]
configuration of $^{42}$Sc. On the contrary, the $[440]1/2(r=+i)$ and $[550]1/2(r=+i)$ 
orbitals have the largest and most dense density clusters in the polar regions of the 
nucleus (see middle of right column in Fig.\ \ref{fig-molecule}); these two orbitals are
occupied in the MD [421,421] configuration. Thus, particle hole excitations 
$[202]5/2$+$[321]3/2$$\rightarrow$$[440]1/2(r=+i)$+$[550]1/2(r=+i)$ performed
both in proton and neutron subsystems  move the matter from the neck (equatorial) region 
into the polar regions of the  nucleus and form the nuclear molecule.

   There are other examples of nuclear molecules in the region of interest.  These are the 
MD configurations [31,31] and [41,41] in $^{36}$Ar (see Fig.\ \ref{fig-chart}) and  MD [42,42] 
configuration in $^{40}$Ca (see Ref.\ \cite{RA.16}). Another well known case of molecular 
structure in this mass region is the superdeformed  configuration [2,2] in $^{32}$S
\cite{MKKRHT.06,MH.04,RA.16}; according to Refs.\ \cite{MKKRHT.06,MH.04} the wavefunction 
of this configuration contains significant admixture of the molecular $^{16}$O+$^{16}$O structure.
Similar to above discussed case of $^{42}$Sc, they are created by respective particle-hole
excitations. For example, nuclear molecule in the MD [31,31] configuration of $^{36}$Ar is
created from the HD [4,4] configuration, which has ellipsoidal density distribution [see Fig. 24b 
in Ref.\  \cite{RA.16}], by particle-hole excitations from the 3/2[321] orbital (which has triple 
ring density distribution (see Fig.\ \ref{fig-Nilsson}) into the [440]1/2$(r=-i)$ orbital.

\section{Conclusions}
\label{concl}

   The competition of different types of nuclear shapes (cluster, ellipsoidal and
nuclear molecules) in light $A=12-50$ nuclei and its evolution with mass
number have been analyzed in covariant density functional theory.  The
occupation of the $[NN0]1/2$ single-particle states favors $\alpha$-clusterization 
because of specific features of the nodal structure of their single-particle density 
distributions.  Density clusters formed by these
states are well localized, have spheroidal or somewhat ellipsoidal density 
distribution and are located on the axis of symmetry of the nucleus. On the 
contrary, the $[N, N-1,1]\Omega$ states have doughnut (for $N=1$)  and 
multiple-ring-type (for $N>1$) density distributions. Thus, the occupation
of such states favors ellipsoidal shapes or nuclear molecules but  disfavors 
$\alpha$-clusterization since such density distributions are incompatible with 
$\alpha$ clusters. The abundance/dominance of such states increases with 
particle number and this explains the transition from frequent appearance of 
cluster structures in very light nuclei with $A\sim 12$ to the dominance of 
ellipsoidal shapes/nuclear molecules in the heavier $A>20$ nuclei.

\section{ACKNOWLEDGMENTS}

I would like to express my gratitute to my collaborators, D. Ray and 
H. Abusara, who contributed to the investigations presented here.
This material is based upon work supported by the U.S. Department 
of Energy, Office of Science, Office of Nuclear Physics under Award 
No. DE-SC0013037.

\bibliographystyle{woc.bst}
\bibliography{man-bulg-sub}

\end{document}